\documentclass[preprint2]{aastex}

\usepackage{natbib}

\newcommand{\gl}{\lambda}

\newcommand{\gs}{\sigma}

\newtheorem{theorem}{Theorem}[section]
\newtheorem{lemma}{Lemma}[section]

\newtheorem{prop}{Proposition}[section]
\newtheorem{claim}{Claim}[section]

\newtheorem{definition}{Definition}[section]

\newtheorem{question}{Question}[section]
\newtheorem{coro}{Corollary}[section]

\newcommand{\beq}{\begin{equation}}
\newcommand{\eeq}{\end{equation}}

\newcommand{\bea}{\begin{array}}
\newcommand{\ena}{\end{array}}
\newcommand{\bds}{\begin {itemize}}
\newcommand{\eds}{\end {itemize}}
\newcommand{\bdf}{\begin{definition}}
\newcommand{\blm}{\begin{lemma}}
\newcommand{\edf}{\end{definition}}
\newcommand{\elm}{\end{lemma}}
\newcommand{\bthm}{\begin{theorem}}
\newcommand{\ethm}{\end{theorem}}
\newcommand{\bprp}{\begin{prop}}
\newcommand{\eprp}{\end{prop}}
\newcommand{\bcl}{\begin{claim}}
\newcommand{\ecl}{\end{claim}}
\newcommand{\bcr}{\begin{coro}}
\newcommand{\ecr}{\end{coro}}
\newcommand{\bquest}{\begin{question}}
\newcommand{\equest}{\end{question}}

\newcommand{\larrow}{{\larrow}}

\newcommand{\argmin}{\ensuremath{\mathrm{arg}\min}}
\newcommand{\argmax}{\ensuremath{\mathrm{arg}\max}}

\newcommand{\Vhat}{{\ensuremath{{\hat{V}}}}}

\def\IC{\mathbb C}
\def\IN{\mathbb N}
\def\IZ{\mathbb Z}
\def\IR{\mathbb R}

\def\shat{^{\mathchoice{}{}%
 {\,\,\smash{\hbox{\lower4pt\hbox{$\widehat{\null}$}}}}%
 {\,\smash{\hbox{\lower3pt\hbox{$\hat{\null}$}}}}}}

\def\bSigma{{
      \ooalign{
      \smash{\hskip.4pt\raise.4pt\hbox{$\Sigma$}}\vphantom{}\crcr
      \smash{\hskip.7pt\raise.6pt\hbox{$\Sigma$}}\vphantom{}\crcr
      \smash{\hbox{$\Sigma$}}\vphantom{$\Sigma$}}
      \vphantom{\hbox{$\Sigma$}}
      }}
\def\bTheta{{
      \ooalign{
      \smash{\hskip.5pt\raise.5pt\hbox{$\Theta$}}\vphantom{}\crcr
      \smash{\hskip.0pt\raise.1pt\hbox{$\Theta$}}\vphantom{}\crcr
      \smash{\hbox{$\Theta$}}\vphantom{$\Theta$}}
      \vphantom{\hbox{$\Theta$}}
      }}
\def\bDelta{{
      \ooalign{
      \smash{\hskip.4pt\raise.4pt\hbox{$\Delta$}}\vphantom{}\crcr
      \smash{\hskip.7pt\raise.6pt\hbox{$\Delta$}}\vphantom{}\crcr
      \smash{\hbox{$\Delta$}}\vphantom{$\Delta$}}
      \vphantom{\hbox{$\Delta$}}
      }}

\makeatletter

\def\bordermatrix#1{\begingroup \m@th
  \@tempdima 8.75\p@
  \setbox\z@\vbox{%
    \def\cr{\crcr\noalign{\kern2\p@\global\let\cr\endline}}%
    \ialign{$##$\hfil\kern2\p@\kern\@tempdima&\thinspace\hfil$##$\hfil
      &&\quad\hfil$##$\hfil\crcr
      \omit\strut\hfil\crcr\noalign{\kern-\baselineskip}%
      #1\crcr\omit\strut\cr}}%
  \setbox\tw@\vbox{\unvcopy\z@\global\setbox\@ne\lastbox}%
  \setbox\tw@\hbox{\unhbox\@ne\unskip\global\setbox\@ne\lastbox}%
  \setbox\tw@\hbox{$\kern\wd\@ne\kern-\@tempdima\left[\kern-\wd\@ne
    \global\setbox\@ne\vbox{\box\@ne\kern2\p@}%
    \vcenter{\kern-\ht\@ne\unvbox\z@\kern-\baselineskip}\,\right]$}%
  \null\;\vbox{\kern\ht\@ne\box\tw@}\endgroup}
\makeatother

\newcommand{\DL}{\begin{dashlist}}
\newcommand{\DLE}{\end{dashlist}}

\makeatletter
\def\argmin{\mathop{\operator@font arg\,min}}
\def\argmax{\mathop{\operator@font arg\,max}}
\makeatother

\newcommand{\SI}{\begin{indlist}}
\newcommand{\EI}{\end{indlist}}

\slugcomment{Submitted to ApJL}

\shorttitle{Closed form maximum entropy image formation}
\shortauthors{Leshem}

\begin{document}

\title{Closed form solution of the maximum entropy equations with application to fast radio astronomical image formation}


\author{Amir Leshem\altaffilmark{1}}
\affil{School of Engineering, Bar-Ilan University, Ramat-Gan, 52900, Israel}
\affil{Faculty of EEMCS, Delft University of Technology, Delft, 2628CD, The Netherlands}

\altaffiltext{1}{This research was supported by a grant from the Netherlands Foundations of Science and Technology,
STW 10459}
\begin{abstract}
In this paper we analyze the maximum entropy image deconvolution. We
show that given the Lagrange multiplier a closed form can be obtained for the image parameters.
Using this solution we are able to provide better understanding of some of the known behavior of
the maximum entropy algorithm. The solution also yields a very efficient implementation of the maximum entropy
deconvolution technique used in the AIPS package. It requires the computation of a single dirty image and inversion of an elementary
function per pixel.
\end{abstract}



\section{Introduction}
Radio astronomical imaging using earth rotation synthesis radio telescopes is an ill-posed problem due to the irregular
sub-Nyquist sampling of the Fourier domain. During the last
40 years many deconvolution techniques have been developed to solve this problem. These algorithms are based on models
for the radio image. Among these techniques we find the CLEAN method by \citet{Hogbom74}, The maximum entropy algorithm
(MEM) by \citet{frieden72}, \citet{gull78}, \citet{ables74} and \cite{wernecke77}, extensions of the CLEAN to support
multi-resolution and wavelets by \citet{bhatnagar04}, \citet{cornwell08} and \citet{bhatnagar08}, non-negative least
squares by \cite{briggs95}, parametric based imaging
by \citet{leshem2000a} and \citet{bendavid08} and sparse $L_1$ reconstruction by \citet{levanda08} and \citet{wioux09}.
While there is a major amount of experience in using these algorithms, we still lack a comprehensive
theoretical analysis. This would become a more critical problem for the future generation of radio interferometers
that will be built in the next two decade such as the square kilometer array (SKA), the Low Frequency Array (LOFAR), The
Allen Telescope Array (ATA), the Long Wavelength Array (LWA) and the Atacama Large Millimeter Array (ALMA). These
radio-telescopes will include many more stations, will have significantly increased sensitivity and some of them will operate at much lower frequencies than previous
radio telescopes, and therefore would be more sensitive to modeling and calibration errors.

The maximum entropy image formation technique is one of the two most popular deconvolution techniques.
The maximum entropy principle was first proposed by \citet{Jaynes57}. \citet{jaynes82} provides a good
overview of the philosophy behind the idea. Since  then it has been used in a wide spectrum of imaging problems.
The basic idea behind MEM is the following:  Among all images which are consistent with the measured
data and the noise distribution not all satisfy the positivity demand,i.e., the sky brightness is a positive function.
Consider only those that satisfy the positivity demand. From these select the one that is most likely to have been
created randomly. This idea has also been proposed by \citet{frieden72} for optical images and applied to
radio astronomical imaging in \citet{gull78}. Other approaches based
on the differential entropy have also been proposed \citet{ables74} and
\citet{wernecke77}. An extensive collection of papers discussing the various
methods and aspects of maximum entropy can be found in the various papers in
\citet{roberts84}. \citet{narayan86} provides an overview of various maximum entropy techniques and the use of the
various options for choosing the entropy measure. Interestingly, in that paper, a closed form solution is given for
the noiseless case, but not for the general case.

The approach of \citet{gull78} begins with a prior image and iterates
between maximizing the entropy function and updating the $\chi^2$ fit to the data. The computation of the image based on
a prior image is done analytically, but at each step the model visibilities are updated,
through a two-dimensional Fourier transform.
This type of algorithm is known as a fixed point algorithm, since it is based on iterating a function until it converges
to a fixed point. While it is known that for the maximum entropy, this approach usually converges, it was recognized that
the convergence can be slow \citep[see][]{narayan86}. Hence, improved methods based on Newton method and the Conjugate
Gradient technique have been proposed by \citep{cornwell85, sault90, skilling84}. These methods perform direct optimization of the entropy
function subject to the $\chi^2$ constraint.

In this paper we will provide a closed form solution for the maximum entropy image formation problem.
This solution provides a novel short proof of the uniqueness of the maximum entropy solution and allows
us to provide a theoretical explanation to the failure of the maximum entropy algorithm in cases of strong point sources.
The explicit expressions for the solution allows us to quantify the effect of the free parameters involved in the maximum
entropy algorithm.
Using the closed form solution we will also develop a new technique for solving the maximum entropy deconvolution
with a fixed number of operations per pixel, except an initial step of gridding, deconvolution and computation of
a single dirty image as described in \citet[see][]{taylor99}. We believe that this paper is a significant step forward
in understanding image deconvolution techniques.

\section{The Maximum Entropy algorithm}
\label{maxent_algo}
We begin with a short description of the maximum entropy algorithm. Following the standard convention
in this field \citep{gull78, cornwell85, sault90}
we present the one dimensional case. Similar results for the two dimensional case are also valid (but require more
complicated notation).

\citet{gull78} showed that the maximum entropy solution is given by solving the
following Lagrangian optimization problem:
\beq
\label{opt_mem}
I^{MEM}=\arg\max_{I}-\sum_{\ell} I_{\ell} \log I_{\ell}-\frac{\gl}{2}\chi^2(V),
\eeq
where
\beq
\chi^2(V)=\sum_{k \in A} \frac{1}{\gs^2_k}\left|\Vhat_k-
V^{\hbox{model}}_k \right|^2,
\eeq
and $\gl$ is a Lagrange multiplier for the constraint that the model
based visibilities should match the measured visibilities.
Taking the derivative with respect to $I_\ell$ we obtain that the solution is given by:
\beq
\label{update_I}
\bea{l}
I_\ell= \\
\exp\left\{-1+\gl\sum_{k \in A} \frac{Re\left(\left(\Vhat_k-
V^{\hbox{model}}_k\right) e^{i\frac{2\pi k\ell}{N}}\right)}{\gs^2_k}\right\},
\ena
\eeq
where the complex exponent comes from the Fourier transform relationship between $V^{\hbox{model}}_k$ and the intensity
at direction $\ell$. The basic maximum entropy algorithm now proceeds by choosing an initial image model (typically flat image)
computing the model based visibilities $V^{\hbox{model}}_k)$ on the grid $A$. Using these visibilities a new model
image is computed by equation (\ref{update_I}). New visibilities are computed from the new model and the process is
iterated until convergence.

\section{Closed form solution of the Maximum Entropy equations}
We will now show that there is a much simpler technique to compute the maximum entropy image than the existing techniques.
The discussion will also provide a novel simple proof for the uniqueness of the maximum entropy solution as well
as a very efficient solution.
The solution depends on the Fourier inversion nature of the interferometric measurement equation, together with the
common practice of gridding and interpolating the data to obtain simple fast Fourier transform between visibility domain
and image domain. This approximation was used by \citet{cornwell85} where the $\chi^2$ approximation was computed in
the image domain using the gridded data. This will enable us to show that the unique maximum entropy function can be
computed independently for each pixel, given a first step of data interpolation in the visibility domain. To that end
assume that
$I=\left<I_\ell:\ell \in L \right>$ is a fixed point of the maximum entropy iteration (\ref{update_I}). Therefore, the
visibilities $V^{\hbox{model}}_k$ are just Fourier transform of
$I$ sampled at the points $k \in A$, where $A$ is the rectangular grid. To simplify the discussion we will assume that
the measurement noise variance $\gs^2_k=\gs^2$ is identical for all visibilities (an assumption
that follows from standard noise calibration practice in radio interferometry) and that a uniform weighting is used.
(we will comment in the end on other weighting schemes).
The right hand side of (\ref{update_I}) is just a Discrete Fourier Transform of the difference between the measured visibilities and the modeled visibilities.
Hence the right hand side is actually the difference between the dirty image $I^D_\ell$ based on the gridded visibilities
and the model image at the point $I_\ell$. Hence the fixed point of the maximum entropy procedure must satisfy
for each pixel $I_\ell$:
\beq
\label{update_I_fixed}
I_\ell=\exp\left\{-1+\frac{\gl}{\gs^2}N \left(I^D_\ell-I_\ell\right)\right\}
\eeq
Alternatively this expression can be derived using Parseval identity, and deriving the $\chi^2$ term with respect
to $I_\ell$ in the image domain.
Further simplification yields:
\beq
\log I_\ell + \frac{N \gl}{\gs^2}I_\ell = \frac{N \gl}{\gs^2}I^D_\ell-1
\eeq
Taking exponent of the two sides yields
\beq
\label{mem_solution}
I_\ell \exp\left\{\frac{N \gl}{\gs^2}I_\ell\right\}=\exp\left\{\frac{N \gl}{\gs^2}I^D_\ell-1\right\}
\eeq
The function $f_a(x)=xe^{ax}$ is monotonically increasing for $a>0$ and all $x$, non negative for all positive values
of $x$ and negative for negative values of $x$. Since the right hand side of (\ref{mem_solution}) is always positive
every pixel in the reconstructed image is positive. By monotonicity of $f_a(x)$ we obtain that $f$ is invertible and the
fixed point is uniquely determined given the Lagrange multiplier.
The solution is now given by:
\beq
\label{mem_closed_solution}
I_\ell=f_a^{-1}\left( \exp\left\{\frac{N\gl}{\gs^2}I^D_\ell-1\right\} \right).
\eeq
where $a=N \gl /\gs^2$.
Equation (\ref{mem_closed_solution}) is very satisfying. It provides us with the desired understanding of the
way the maximum entropy solution manipulates the gridded data. For each pixel of the dirty image $I^D_{\ell}$ we obtain
that if $1<<N \gl I^D_{\ell}/\gs^2$ then
\beq
\frac{\gs^2}{N \gl I^D_{\ell}} \log I_\ell + I_\ell \approx I^D_{\ell}.
\eeq
Hence whenever the dirty image is sufficiently strong we give preference to the dirty image. If on the other hand, the dirty image is
weak, then there is higher risk of obtaining a sidelobe, so the prior is more significant, and sidelobes are more significantly
suppressed. Note that if we have a very strong point source in the field of view, its sidelobes are also likely to
be stronger than other emission in the image, and therefore the algorithm will not be able to suppress these sidelobes.
This provides full explanation for the problems of maximum entropy deconvolution as described by Briggs et al.
\citep[see][chapter 8]{taylor99}.

\section{Extensions}

\begin{table}
\centering
\caption{Exact maximum entropy algorithm}
\begin{tabular}{||l||}
\hline
\hline
{\bf Initialization:}  \\
Choose a prior image $F_\ell$.\\
Choose $\gl$\\
Grid the data to obtain $\Vhat_k: k\in A$ \\
\quad where $A$ is a uniform rectangular grid.\\
Compute the dirty image $I^D$. \\
\hline
Image deconvolution: \\
For each pixel $\ell$ \\
\quad Compute the closed form solution based on (\ref{mem_solution_ref}).\\
\quad When $I^D_{\ell}$ is negative: \\
\quad \quad choose the solution with maximum entropy. \\
End.\\
Compute fit to data. \\
If fit is poor, increase $\gl$ and redo deconvolution.\\
\hline
\hline
\end{tabular}
\label{exact_mem}
\end{table}
The $\chi^2$ constraint is in many cases insufficient, and either a reference image or a constraint on the total flux is
added in order to provide better fit is added. Our solution naturally extends to these cases. We will demonstrate it in
the first case. Assume that we have a reference image. The problem (\ref{opt_mem}) can be reformulated as
\beq
\label{opt_mem_ref}
\bea{ll}
I=\arg\max_{I}&-\sum_{\ell} I_{\ell} \log \frac{I_{\ell}}{F_\ell} \\
              &-\frac{\gl}{2}\sum_{k \in A} \frac{1}{\gs^2_k}\left|\Vhat_k-
V^{\hbox{model}}_k \right|^2,
\ena
\eeq
where $F_\ell$ is a reference image (possibly low resolution image we already have). The ratio is taken from
the expression for the relative entropy or the Kullback-Leibler divergence.
Taking the derivative with respect to $I_\ell$ and assuming again that $\gs^2_k=\gs^2$ results in the following solution
\beq
\label{update_I_ref}
I_\ell=\exp\left\{-1+F_\ell+\frac{N\gl}{\gs^2}\left(I^D_\ell-I_\ell\right)
\right\}.
\eeq
Hence the solution is given by
\beq
\label{mem_solution_ref}
I_\ell = f_a^{-1}\left(\exp\left\{\frac{N \gl}{\gs^2}I^D_\ell-1+F_\ell\right\}\right).
\eeq
where $a=N \gl I^D_\ell/\gs^2$ and $f^{-1}_a(x)$ is computed as before.

Now that we have explicit expression of the reconstructed image given $\gl$ we can use either a bisection or Newton
type technique to solve for the Lagrange multiplier that provides sufficiently good fit for the data. We will not provide
explicit expressions, due to lack of space. The algorithm is described in Table \ref{exact_mem}.

Finally we comment that when non uniform weighting is used, closed form solution is harder, as already noted by
\cite{cornwell85}. We adopt their solution and replace the $\chi^2$ term by the same image domain expression. The only
thing that changes is the computation of the weighted dirty image. Since this technique has been used in
the last three decades it is safe to believe that the impact on the ME algorithm is small.

\section{Conclusions}
In this paper we have demonstrated that the maximum entropy algorithm can be solved analytically. The solution is
a two steps approach: Gridding of the visibilites into a rectangular grid and solution of the gridded maximum entropy
problem in the image domain. We showed that the solution can be computed analytically requiring only an inversion of an
elementary function for each pixel. The solution provides better theoretical understanding of the maximum entropy algorithm as well as significant
acceleration of the algorithm implemented in existing radio astronomy packages, by reducing the number of
Fourier transforms from 20-100 to a single dirty image computation. This enhancement will allow much faster maximum entropy
deconvolution for very large images. We do not provide simulated comparison, since by our results,
the proposed algorithm is identical to the maximum entropy algorithm as implemented in the VM and VTESS tasks in the
AIPS package.
\bibliographystyle{abbrvnat}

\begin{thebibliography}{21}
\providecommand{\natexlab}[1]{#1}
\providecommand{\url}[1]{\texttt{#1}}
\expandafter\ifx\csname urlstyle\endcsname\relax
  \providecommand{\doi}[1]{doi: #1}\else
  \providecommand{\doi}{doi: \begingroup \urlstyle{rm}\Url}\fi

\bibitem[Ables(1974)]{ables74}
J.~Ables.
\newblock Maximum entropy spectral analysis.
\newblock \emph{Astronomy and Astrophysics Supp.}, 15:\penalty0 383--393, 1974.

\bibitem[Ben-David and Leshem(2008)]{bendavid08}
C.~Ben-David and A.~Leshem.
\newblock Parametric high resolution techniques for radio astronomical imaging.
\newblock \emph{Selected Topics in Signal Processing, IEEE Journal of},
  2\penalty0 (5):\penalty0 670--684, Oct. 2008.

\bibitem[Bhatnager and Cornwell(2004)]{bhatnagar04}
S.~Bhatnager and T.~Cornwell.
\newblock Adaptive scale sensitive deconvolution of interferometric images {I}.
  {A}daptive scale pixel (asp) decomposition.
\newblock \emph{Astronomy and Astrophysics}, 426:\penalty0 747--754, 2004.

\bibitem[Briggs(1995)]{briggs95}
D.~S. Briggs.
\newblock \emph{High fidelity deconvolution of moderately resolved sources}.
\newblock PhD thesis, The new Mexico Institute of Mining and Technology,
  Socorro, New Mexico, 1995.

\bibitem[Cornwell(2008)]{cornwell08}
T.~Cornwell.
\newblock Multiscale clean deconvolution of radio synthesis images.
\newblock \emph{Selected Topics in Signal Processing, IEEE Journal of},
  2\penalty0 (5):\penalty0 793--801, Oct. 2008.

\bibitem[Cornwell and Evans(1985)]{cornwell85}
T.~Cornwell and K.~Evans.
\newblock A simple maximum entropy deconvolution algorithm.
\newblock \emph{Astronomy and Astrophysics}, 143:\penalty0 77--83, 1985.

\bibitem[Cornwell et~al.(2008)Cornwell, Golap, and Bhatnagar]{bhatnagar08}
T.~Cornwell, K.~Golap, and S.~Bhatnagar.
\newblock The noncoplanar baselines effect in radio interferometry: The
  w-projection algorithm.
\newblock \emph{Selected Topics in Signal Processing, IEEE Journal of},
  2\penalty0 (5):\penalty0 647--657, Oct. 2008.

\bibitem[Frieden(1972)]{frieden72}
B.~Frieden.
\newblock Restoring with maximum likelihood and maximum entropy.
\newblock \emph{Journal of the Optical Society of America}, 62:\penalty0
  511--518, 1972.

\bibitem[Gull and Daniell(1978)]{gull78}
S.~Gull and G.~Daniell.
\newblock Image reconstruction from incomplete and noisy data.
\newblock \emph{Nature}, 272:\penalty0 686--690, 1978.

\bibitem[Hogbom(1974)]{Hogbom74}
J.~Hogbom.
\newblock Aperture synthesis with non-regular distribution of interferometer
  baselines.
\newblock \emph{Astronomy and Astrophysics Supp.}, 15:\penalty0 417--426, 1974.

\bibitem[Jaynes(1957)]{Jaynes57}
E.~Jaynes.
\newblock \emph{Physics Review}, 106:\penalty0 620, 1957.

\bibitem[Jaynes(1982)]{jaynes82}
E.~Jaynes.
\newblock On the rational of maximum entropy methods.
\newblock \emph{Proceedings of the IEEE}, 70:\penalty0 939--952, 1982.

\bibitem[Leshem and van~der Veen(2000)]{leshem2000a}
A.~Leshem and A.~van~der Veen.
\newblock Radio-astronomical imaging in the presence of strong radio
  interference.
\newblock \emph{IEEE Trans. on Information Theory, Special issue on information
  theoretic imaging}, pages 1730--1747, 2000.

\bibitem[Levanda and Leshem(2008)]{levanda08}
R.~Levanda and A.~Leshem.
\newblock Radio astronomical image formation using sparse reconstruction
  techniques.
\newblock pages 716--720, Dec. 2008.

\bibitem[Narayan and Nityananda(1986)]{narayan86}
R.~Narayan and R.~Nityananda.
\newblock Maximum entropy image restoration in astronomy.
\newblock \emph{Annual review of of Astronomy and Astrophysics}, 24:\penalty0
  127--170, 1986.

\bibitem[Roberts(1984)]{roberts84}
J.~Roberts, editor.
\newblock \emph{Indirect imaging}.
\newblock Cambridge university press, 1984.

\bibitem[Sault(1990)]{sault90}
R.~Sault.
\newblock A modification of the {C}ornwell and {E}vans maximum entropy
  algorithm.
\newblock \emph{The Astrophysical Journal}, 354:\penalty0 L61--63, 1990.

\bibitem[Skilling and Bryan(1984)]{skilling84}
J.~Skilling and R.~Bryan.
\newblock maximum entropy image restoration algorithm.
\newblock \emph{Monthly Notices of the Royal Astronomical Society},
  211:\penalty0 111--124, 1984.

\bibitem[Taylor et~al.(1999)Taylor, Carilli, and Perley]{taylor99}
G.~Taylor, C.~Carilli, and R.~Perley.
\newblock \emph{Synthesis Imaging in Radio-Astronomy}.
\newblock Astronomical Society of the Pacific, 1999.

\bibitem[Wernecke(1977)]{wernecke77}
S.~Wernecke.
\newblock Two dimensional maximum entropy reconstruction of radio brightness.
\newblock \emph{Radio Science}, 12:\penalty0 831--844, 1977.

\bibitem[Wiaux et~al.(Submitted 2009)Wiaux, Jacques, Puy, Scaife, and
  Vandergheynst]{wioux09}
Y.~Wiaux, L.~Jacques, G.~Puy, A.~Scaife, and P.~Vandergheynst.
\newblock Compressed sensing imaging techniques for radio interferometry.
\newblock Submitted 2009.

\end{thebibliography}

\end{document}